%% file: ipdps-conf.tex
\ifdefined\singlecol
\documentclass[11pt, conference, compsocconf, onecolumn]{IEEEtran}
\else
\documentclass[10pt, conference, compsocconf]{IEEEtran}
\fi
\include{preamble}

\ifCLASSINFOpdf
\else
\fi
\usepackage[caption=false,font=footnotesize]{subfig}


\begin{document}

%\title{Composable notations for distributed parallel computations}
\title{Scalable data abstractions for distributed parallel computations}


\author{\IEEEauthorblockN{James~Hanlon, Simon~J~Hollis and David May}
\IEEEauthorblockA{Department of Computer Science\\
University of Bristol\\
Bristol, UK\\
\{hanlon, simon, dave\}@cs.bris.ac.uk}
}

\maketitle

\begin{abstract}

The ability to express a program as a hierarchical composition of parts is an
essential tool in managing the complexity of software and a key abstraction
this provides is to separate the representation of data from the computation. 
Many current parallel programming models use a shared memory model to provide
data abstraction but this doesn't scale well with large numbers of cores due to
non-determinism and access latency.
This paper proposes a simple programming model that allows scalable parallel
programs to be expressed with distributed representations of data and it
provides the programmer with the flexibility to employ shared or distributed
styles of data-parallelism where applicable.
It is capable of an efficient implementation, and with the provision of a small
set of primitive capabilities in the hardware, it can be compiled to operate
directly on the hardware, in the same way stack-based allocation operates for
subroutines in sequential machines.

\end{abstract}


% \fi
\IEEEpeerreviewmaketitle

\ifdefined\singlecol
  \setlength\parindent{0mm}
  \setlength\parskip{10pt}
\fi

\section{Introduction}
% Composing paradigms and (distributed) representations of data
% General principles applied to expressing general purpose programs.

% Importance of abstraction
When developing a program of any complexity, the ability to express it in terms
of a simpler set of components is essential. A component presents a simple
interface that allows its implementation to be considered independently, and
when combined with other components, the internal details can be ignored and
its functionality treated in an abstract way. This allows a program to be
constructed using modules, ranging from small functions to libraries, and for
any component to be substituted with another that adheres to the same
interface.
The importance of abstraction as a tool in computer programming was recognised
by Turing in the 1940s~\cite{Turing46} and was formalised in the 1970s by the
\emph{structured programming} methodology~\cite{Dahl72}. This aimed to improve
the quality of programs and productivity of programmers through judicious use
of \emph{hierarchical structuring} and \emph{subroutines}.  These principles
have been foundational for modern sequential programming languages.

% Separating data and shared memory
A key issue with composability is separating the representation of data from
the structure of a computation. 
Mainstream CPU and general-purpose GPU (GPGPU) parallel programming models are
based on a shared memory model, where data are globally accessible. This is the
form of \emph{parallel random access machine} (PRAM)~\cite{Fortune78} and the
related \emph{bulk-synchronous parallel} (BSP)~\cite{Valiant90a} model. 
Shared memory parallelism allows sequential approaches to data abstraction and
conventional data structures to be employed, but it does not scale well with
large numbers of cores. 
Access latency can vary significantly and unpredictably due to the physical
distribution of data across a machine. This makes it difficult to exploit
locality, which is essential for scaling a computation, and poses problems for
barriers which are delayed by the slowest participant. Additionally, when
accesses are made to shared data they can incur latency from collisions, and
when they are updating it, behaviour can become non-deterministic.

% Mainstream: Multi-threading/SMPs (resource allocation) 
There are a number of issues related to the implementation of a shared memory
system that pose further problems for this type of data abstraction.
Mainstream parallel processors take the form of symmetric multi-processors
(SMPs) and these have brought about a number of parallel programming approaches
such as the Cilk~\cite{Cilk95} language, OpenMP~\cite{OpenMP} and Intel's
Threaded Building Blocks (TBB)~\cite{IntelTBB}. These  employ a
\emph{multi-threaded} execution model where a number of threads are managed by
a scheduler, but problems can arise with programs that combine parallel
components.
Performance can be affected significantly by threads competing for execution,
causing unnecessary context-switches, and idling within a component due to a
load imbalance, causing under utilisation. The effects of this are dependent on
combinations of program components and result in unpredictable execution time,
exacerbating non-deterministic behaviour.
%
% Emerging mainstream: heterogeneous
OpenCL~\cite{OpenCL} is emerging into the mainstream and is designed to support
the programming of heterogeneous systems. These are typically comprised of CPUs
and GPGPUs. It uses a shared memory model but exposes distinct address spaces
and in order to compose components operating in different ones, variables must
be explicitly transferred between them~\cite{Gaster12}. 

% Mainstream models and their problems (SMP CPU)
%Characteristics of current CPU and GPU architectures pose significant problems
%for composability, making it difficult to construct modular reusable parallel
%components such as libraries.
%%
%% Reproducable behaviour
%The most significant issue is allocating processing resources to components in
%such a way that their behaviour is predictable. This is essential for
%components to be treated abstractly according to their interface. Without it,
%their internal details have to be included in the reasoning about the whole
%system~\cite{Hansen02}. 
%%
% Heterogeneous
%Another problem for composability is the use of machine-orientated features.
%With these, programs are not understandable in terms of the concepts of a
%language and thus components cannot be composed only in terms of the language.
%

% Distributed representation of data
Parallelism is now the primary means of sustaining growth in computational
performance~\cite{FutureCompPerf} and the shared memory model will continue to
be useful. However, it looks certain that future systems will involve large
numbers of processors and it will not be effective in delivering performance on
them.
Therefore, it is necessary that parallel programming models, as well as
supporting shared memory approaches, also support composable representations of
distributed data.

% Proposed solution
This paper proposes a simple distributed programming model that builds on the
approach of the occam programming language~\cite{Occam83} with notations to
control the distribution of parallelism and a \emph{server} construct that is
active only in response to requests. 
Arrays of servers can be combined to construct distributed data structures,
independently from the computational aspects of a program, providing access for
shared or distributed styles of data-parallelism.  This gives the programmer
flexibility to employ the most appropriate data representation for the purposes
of the program and scalability. Server-based data structures can be composed
with similar scoping rules to conventional variable declarations to simplify
the task of building scalable programs by allowing them to be composed in a
modular way.
With the provision of a small set of primitive capabilities in the hardware,
the model can be compiled with a fixed allocation of processors. This is so it
can operate efficiently and directly on the hardware, without the use of
dynamic allocation mechanisms. The idea is similar to stack-based allocation
for subroutines in sequential machines. 

% Contributions
The following specific contributions are made:
\ben

\item A server construct that can be used to express composable
  representations of distributed data structures with arrays of server
  processes, for both shared memory and message passing distributed memory
  style parallel computations.

\item An efficient implementation of distributed parallelism
  based on a compile-time allocation of processors.

\item An implementation of server processes that allows many-to-one client
  connections to be established efficiently and without deadlock.

\item Demonstration of the proposed notations with three example programs that
  are characteristic of general-purpose applications and employ different
  styles of parallelism. 

\een

% Overview
The rest of this paper is organised as follows.
\sect{related} overviews related work;
\sect{proposal} presents the proposed programming model and notations in terms
of a conceptual machine model;
\sect{compilation} describes the requirements of a target architecture and the
compilation scheme for it;
\sect{examples} discusses how several example programs that require distinct
styles of parallelism can be expressed in the model;
%
%\sect{discussion} discusses potential ways in which the proposal could be
%improved or extended;
%
\sect{conclusion} concludes.

\section{Related work\label{sec:related}}

% Distributed
Distributed memory architectures are most common in high performance computing
(HPC) systems and the Message Passing Interface (MPI)~\cite{MPI2.2} is the
standard programming approach. 
%
% MPI (local view)
MPI provides features for the construction of modular components such as
libraries~\cite{Skjellum94} with features to name groups of processes and
provide scoping for operations within them, but it does not allow a separation
of data because of its SPMD (single program multiple data) model.
The success of MPI can be attributed to its simple compilation and execution
model, which provides predictable execution that allows programmers to make
efficient use of a machine. Other more dynamic languages push resource
allocation and management into runtime components that require significant
overheads in execution time and storage, and result in less predictable
execution of program components.

% MPI dyamic process creation
Dynamic process creation was introduced in MPI-2, and in particular, a server
construct, similar to the proposal in this paper, was introduced to address the
need to support groups of reactive processes that accept connections from other
groups~\cite[Sect.~10.4]{MPI2.2}.  The problem with this is that the location
of processes is not known at compile-time. To quote the specification directly
\emph{Almost all of the complexity in MPI client/server routines addresses the
question ``how does the client find out how to contact the server?''}.
This issue also lies at the heart this work, but the solution is simplified by
the choice of notations, the restrictions placed on them and support required
in the architecture.

% PGAS (global view)
Partitioned global address space (PGAS) languages such as UPC~\cite{UPC99},
Chapel~\cite{Chapel07} and X10~\cite{X1005} are based on globally accessible
variables that are divided into logical segments to provide a clean composition
of distributed data and computation. These segments have affinity with
particular processes to provide a notion of locality for fast memory accesses,
and global accesses are compiled into message passing communications. 
% Efficiency concerns
These languages include a range of distributed data types with high level
notations for operating on them. Static distributions can be compiled into
message passing programs, although it is not yet clear how efficient they are
compared to manually crafted MPI equivalents, and as yet PGAS languages have
not had widespread adoption.

% Charm
Charm~\cite{Charm95} is another HPC-orientated language but takes a different
approach. Parallelism is expressed with arrays of objects and communication is
performed with remote method calls. A runtime system is responsible for
dynamically mapping objects onto processors and scheduling communication.  As
is the case with dynamic processes in MPI, this requires all communications to
be directed through proxy processes aware of object locations. Although Charm
encourages modular development, it does not directly support composable
representations of distributed data.

% Occam/XC TODO? PBH paradigms
Occam~\cite{Occam83} and its descendant XC~\cite{XMOSLang} are message passing
languages for distributed memory architectures.  Predicable execution is a key
principle of them and this is achieved primarily with a compile-time allocation
of memory by prohibiting recursion and dynamically sized arrays.
Implementations require the allocation of processors to be specified
statically in a mapping file and program components cannot employ distributed
parallelism internally.
Developments as part of the occam 3 specification introduced the concept of a
server component~\cite[Chapt.~13]{Occam3Ref}. The proposed notation builds on
this with a distributed execution model and relaxed communication constraints.

\section{Proposal\label{sec:proposal}}

\subsection{Architectural model}

% Model
The proposed programming model is based on a simple conceptual architecture
where, to a first order approximation, there is an infinite array of
processors.  Each processor has a relatively small private memory, but the
ability to communicate with any other processor via a network in a constant
amount of time, independent of the processor locations.  This is an idealised
view held by the programmer to simplify programming.

% Realistic machine (emulation)
A realistic parallel machine can provide a good approximation to this with a
fixed number of processors and a logarithmic-diameter, high-capacity network
such as a Clos/fat tree~\cite{Leiserson85} or hypercube~\cite{Valiant90}.
Networks such as meshes do not provide these properties and programs must be
carefully mapped to preserve locality to obtain good performance.

% RAM model analogy
This model is analogous to the \emph{random access machine} (RAM)
model of computation~\cite{Cook72} which models the essential aspects of a
conventional sequential computer. It consists of a program that operates on an
infinite capacity memory where accesses take a constant amount of time,
independent of the address.  In practical sequential computers, memory size
is limited and access incurs a latency related to capacity, also by a
logarithmic scaling.

\subsection{Notations}

The following is an informal description of the proposed language notations. An
imperative block-structured syntax is used and the basic features of this are
based on the occam programming language~\cite{Occam83}. It includes sequential
and parallel composition, replication and channel-based communication and
provides a platform for the main contributions of this paper: 
notations to express local and distributed parallelism and a \emph{server}
construct.
\emph{Local} parallelism relates to concurrent threads that access a
shared memory and \emph{distributed} relates to distinct memories.
Diagrams are included throughout to provide an intuition for the programming
model and behaviour of the notations in isolation and in composition.

\subsubsection{Composition}

A program is built as a hierarchical collection of \emph{processes} that can be
composed in sequence or in parallel. \emph{Sequential composition} is denoted
by the `\ttt{;}' separator and causes a set of processes to be executed one
after another. If $P$, $Q$ and $R$ are processes, then the process 
\code{$P$ ; $Q$ ; $R$} 
is executed by running $P$, $Q$ and then $R$.  Sequential composition can be
\emph{replicated} to produce a number of similar processes executed in
sequence. If $P(i)$ is a process, then the process
\code{\w{seq} i=1 \w{for} $n$ \w{do} $P$(i)}
is equivalent when $n=4$ to \code{$P(0)$ ; $P(1)$ ; $P(2)$ ; $P(3)$.}
%
%\inputtikz{figures/seq-composition}
%
\emph{Parallel composition} causes the component processes to start
simultaneously and the execution can be directed to occur locally or
distributed over an array of processors.  Local parallel execution is denoted
by the `\ttt{|}' separator. The process
\code{$P$ | $Q$ | $R$.}
causes the component processes $P$, $Q$ and $R$ to start simultaneously on a
processor $p_k$, where $k$ is the identifier (ID) of a processor, and it
terminates when all component processes have terminated.
%

%\tikzset{/tikz/external/export=true}%
%\tikzsetnextfilename{#1}%
%\tikzpicturedependsonfile{#1.tex}%
%\tikzsetnextfilename{#1}%
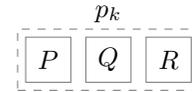
\begin{center}%
\input{figures/par-composition-local}%
\end{center}%
%\tikzset{/tikz/external/export=false}

%
Distributed parallel composition is denoted by the `\ttt{\&}' separator.
The process
\code{$P$ \& $Q$ \& $R$} 
is equivalent to the above local composition, except that $P$, $Q$
and $R$ start simultaneously on different processors $p_k, p_{k+1}, p_{k+2}$.
%

%\tikzset{/tikz/external/export=true}%
%\tikzsetnextfilename{#1}%
%\tikzpicturedependsonfile{#1.tex}%
%\tikzsetnextfilename{#1}%
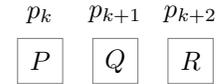
\begin{center}%
\input{figures/par-composition}%
\end{center}%
%\tikzset{/tikz/external/export=false}

%
Distributed composition can be replicated to produce a number of similar
parallel processes and can be thought of as declaring a \emph{process array}.
The process 
\code{\w{par} i=1 \w{for} $n$ \w{do} $P$(i)}
is equivalent when $n=4$ to
\code{$P(0)$ \& $P(1)$ \& $P(2)$ \& $P(3)$.}

%causes $n$ instances of the process $P$ to start simultaneously on $n$
%different processors, $p_k, p_{k+1}, \cdots, p_{k+n-1}$.
%
%\inputtikz{figures/par-replication}
%
%It
%terminates when all component processes have terminated. 

%\subsubsection{Replication}
%
%Both sequential and parallel composition can be \emph{replicated} to trigger
%the execution of an \emph{array} of processes, and these processes can be
%parametrised by indices defined in the replicator. 
%
%and each time, the value of the
%\emph{replicator index} $i$ is incremented by one and $P$ sees that unique
%value of $i$. 
%%
%%\inputtikz{figures/seq-replication}
%%
%The replicator index cannot be changed by any of the component processes to
%enforce this. 
%%
%Similarly, for parallel replication, the process
%%
%\code{\w{par} i=1 \w{for} $n$ \w{do} $P$(i)}
%%
%causes $n$ instances of the process $P$ to start simultaneously on $n$
%different processors, $p_k, p_{k+1}, \cdots, p_{k+n-1}$.  The process
%terminates when all component processes have terminated. This can be thought of
%as declaring a \emph{process array}.
%%
%\inputtikz{figures/par-replication}
%%
%For $n=4$ the above process is equivalent to the explicit distributed
%composition
%%
%\code{$P(0)$ \& $P(1)$ \& $P(2)$ \& $P(3)$.}
%%
%Unlike sequential replication, the \emph{count}, in this case $n$, must be a
%constant to allow a compile-time allocation of processors (explained in
%\sect{compilation}).  This is similar to constant-value lengths of array
%declarations in C for compile-time stack allocation.

% Allocation and layering
Processes in distributed composition are allocated on consecutively numbered
processors as this simplifies the task of establishing communication channels
with them because they can addressed with a base and offset.
This property of the notation also allows correspondences to be
established between different arrays.
%
% Corresponding process arrays
For example, replication, combined with local composition can be used to
\emph{layer} arrays of parallel processes on the same array of processors. The
process
\begin{code*}
\w{par} i=1 \w{for} \(m\) \w{do} \(P\)(i) | 
\w{par} i=1 \w{for} \(n\) \w{do} \(Q\)(i)
\end{code*}
for $m=n$ causes each processor $p_k,p_{k+1},\cdots,p_{k+n-1}$ to execute
$P(x)$ and $Q(x)$ for some $x$.
%

%\tikzset{/tikz/external/export=true}%
%\tikzsetnextfilename{#1}%
%\tikzpicturedependsonfile{#1.tex}%
%\tikzsetnextfilename{#1}%
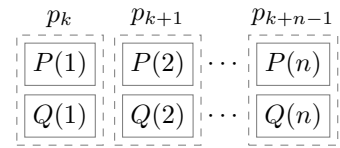
\begin{center}%
\input{figures/par-replication-layered}%
\end{center}%
%\tikzset{/tikz/external/export=false}

%
The result of this is a direct correspondence between $P$ and $Q$
with the same index and any communication between them will be performed
locally. For $m\neq n$, one array will be larger and allocated over more
processors.
In contrast, the distributed composition of the same replicators
\begin{code*}
\w{par} i=1 \w{for} \(m\) \w{do} \(P\)(i) \& 
\w{par} i=1 \w{for} \(n\) \w{do} \(Q\)(i)
\end{code*}
allocates both process arrays on disjoint sets of processors
%

%\tikzset{/tikz/external/export=true}%
%\tikzsetnextfilename{#1}%
%\tikzpicturedependsonfile{#1.tex}%
%\tikzsetnextfilename{#1}%
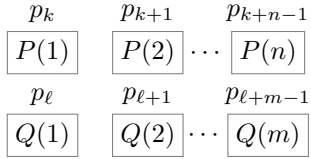
\begin{center}%
\input{figures/par-replication-disjoint}%
\end{center}%
%\tikzset{/tikz/external/export=false}

%
for $\ell \geq k+n$.

\subsubsection{Servers\label{sec:servers}}

% Servers to separate data and different data-parallelism
The server notation provides a simple way of separating a representation of
data from the computations which act on it and can be used in conjunction with
replicators to implement distributed structures that can be accessed
concurrently.
Furthermore, it allows both shared and distributed memory style parallelism to
be expressed in a similar way. This is a significant capability as it allows a
programmer to move easily between them. 

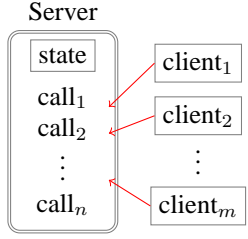
\begin{figure}[t]

%\tikzset{/tikz/external/export=true}%
%\tikzsetnextfilename{#1}%
%\tikzpicturedependsonfile{#1.tex}%
%\tikzsetnextfilename{#1}%
\begin{center}%
\input{figures/server}%
\end{center}%
%\tikzset{/tikz/external/export=false}

\caption{A server process, serving a set of clients.}

\label{fig:server}
\end{figure}

% Description
A \emph{server} is a special kind of process that is only active in response to
\emph{clients}. The interface to a server is a set of \emph{calls}, which
behave in the same way as conventional procedure calls, except the parameters
and results are transferred to and from the server so that execution of the
call occurs at the server. \fig{server} illustrates a single server with a set
of clients.
This mechanism is known generally as a \emph{remote procedure call}
(RPC)~\cite{Birrell84} and is attractive because it provides clean semantics,
hiding the underlying communication, and provides the ability to move easily
between the local and remote forms of a call.

% Server definitions
A server definition specifies a set of potential calls and provides responses
to them.  Its only action while running is to repeatedly serve calls and it
terminates when its scope terminates.  Local state can be initialised by a
special initialisation process and a corresponding termination process can be
used to finalise the server upon termination.
In \emph{object-orientated} programming, this relates directly to the concept
of an object with a constructor that takes an initial value and methods that
operate on the private attributes.

% Store example
As an example, \proc{store} defines a server to provide access to an array.
When it initialises, each element of the array is set to an initial value,
specified as a parameter (\ttt{init}), and when the server is running, calls
can be made to \ttt{read} or \ttt{write} to specific locations. 
\begin{Process}
\begin{code*}
\w{server} Store(\w{val} init)
  \w{interface}(
    \w{call} read(\w{val} i, \w{var} v), 
    \w{call} write(\w{val} i, \w{val} v)) \w{to}
\{ \w{var} data[N];
  \w{inital}
  \{ \w{var} i;
    \w{seq} i=0 \w{for} N \w{do}
      data[i] := init
  \}
  \w{accept}
  \{ read ? (\w{val} i, \w{var} v)
      v := data[i]
    write ? (\w{val} i, \w{val} v)
      data[i] := v
  \}
  \w{final} \{\}
\}
\end{code*}
\label{proc:store}
\end{Process}

The following \emph{specifies} an instance of the \ttt{Store} server with the
name \ttt{s}, for use with an anonymous client process that executes in
parallel and makes calls to write to each store location.
\begin{code*}
\w{server} s \w{is} Store(0) \&
\w{seq} i=0 \w{for} \(n\) \w{do} s.write(i, i)
\end{code*}

% Server arrays
Servers can be replicated with a similar notation to a conventional array
declaration. For example the \emph{server array}
\code{\w{server} s \w{is} Store(0)[\(n\)]\\$\cdots$}
creates $n$ instances of the \ttt{store} server, with each initialised by the
same parameters, in this case 0.  A call to a particular server is made by
specifying a server with an array subscript such as \ttt{s[0]}.
%

%\tikzset{/tikz/external/export=true}%
%\tikzsetnextfilename{#1}%
%\tikzpicturedependsonfile{#1.tex}%
%\tikzsetnextfilename{#1}%
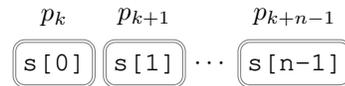
\begin{center}%
\input{figures/server-replicated}%
\end{center}%
%\tikzset{/tikz/external/export=false}

\subsection{Expressing data-parallelism}

% Expressing data-parallelism
With the proposed notations for controlling distribution and creating arrays of
servers that can be accessed by collections of clients, it is possible to
express both shared and distributed memory forms of data-parallel computations.

\subsubsection{Shared memory}

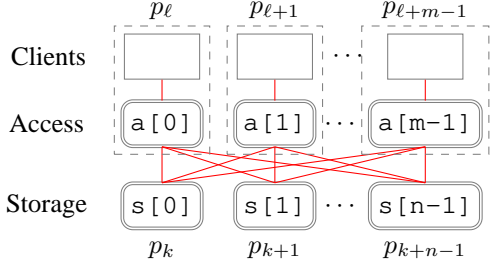
\begin{figure}[t]

%\tikzset{/tikz/external/export=true}%
%\tikzsetnextfilename{#1}%
%\tikzpicturedependsonfile{#1.tex}%
%\tikzsetnextfilename{#1}%
\begin{center}%
\input{figures/server-distributed-shared-memory}%
\end{center}%
%\tikzset{/tikz/external/export=false}

\caption{Illustration of the layout and structure of the shared memory
implementation in \proc{shared-memory}. The storage is distributed over a
disjoint array of processors, hence $\ell \geq k+n$.}
\label{fig:shared-memory}
\end{figure}

% Distributed shared memory
A shared memory, distributed over an array of processors, can be expressed with
two server arrays, one to act as a store and the other to provide an access
abstraction.
For example, \proc{shared-memory} provides an access server (\ttt{Access} which
has the same interface as \ttt{Store}) to each of the $m$ client processes.
\begin{Process}
\begin{code*}
\w{server} s \w{is} Store(0)[\(n\)] &
\{ \w{server} a \w{is} Access(s)[\(m\)] |
  \w{par} i=0 \w{for} \(m\) \w{do}
  \{ \(\cdots\); a[i].write(\(\cdot\), \(\cdot\)); \(\cdots\) \}
\}
\end{code*}
\label{proc:shared-memory}
\end{Process}
The access and client processes are layered over the same processors so
interaction between these is local. Each access server holds a reference to the
array of $n$ storage servers and takes read and write requests from the client and
performs them over this array. \fig{shared-memory} illustrates the layout and
structure of this.

To avoid uneven distribution of accesses and load on particular servers, which
would result in increased access latency, the access servers could select
storage servers by some appropriate hash function.  This is the form of a PRAM
and the memory system of a BSP machine.  For the most general concurrent-read
concurrent-write (CRCW) form of memory, read combining could also be used to
avoid excessive access collisions~\cite{Ranade87}.

\subsubsection{Distributed memory}

% Distributed memory
A distributed representation of data can be expressed in a similar way, without
an access abstraction and with the server and client processes distributed over
the same set of processors. \proc{distributed-memory} is similar to
\proc{shared-memory}, except clients are co-located with a storage
server and access it directly.
\begin{Process}
\begin{code*}
\w{server} s \w{is} Store(0)[\(n\)] |
\w{par} i=0 \w{for} \(n\) \w{do}
\{ \(\cdots\); a[i].write(\(\cdot\), \(\cdot\)); \(\cdots\) \}
\end{code*}
\label{proc:distributed-memory}
\end{Process}
Since there is a local correspondence between servers and clients, this call
will not incur any overhead due to the underlying interconnection network. 

\begin{figure}[t]

%\tikzset{/tikz/external/export=true}%
%\tikzsetnextfilename{#1}%
%\tikzpicturedependsonfile{#1.tex}%
%\tikzsetnextfilename{#1}%
\begin{center}%
\input{figures/server-distributed-memory}%
\end{center}%
%\tikzset{/tikz/external/export=false}

\caption{The layout and structure of the distributed memory
\proc{distributed-memory}.  Servers are situated with clients for fast access.}

\label{fig:distributed-memory}
\end{figure}
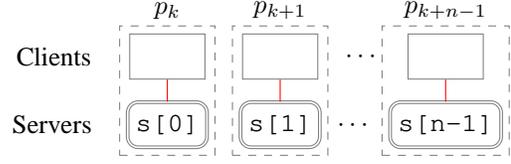

% Employing message passing communication
Data stored with other server or client processes could be accessed with server
calls, but race conditions can arise from concurrent access to shared data and
synchronisation is required to avoid this.
Instead, synchronised message passing communication avoids these issues and is
widely used for scalable algorithms, typically in large systems such as
supercomputers. In general, simple scalable structures such as pipelines, grids
and trees are used~\cite{Hansen93} which are easily expressed in occam and
hence are composable with server-based representations of data. 
This is demonstrated with the matrix multiplication example in
\sect{matrix-multiply}.

% Summary
Shared and distributed memory forms of data-parallelism lend themselves to
different applications and the ability of the proposed programming model to
cleanly support both is significant. It provides the programmer with the
flexibility to employ a notation that best suits a given application.

\section{Compilation\label{sec:compilation}}

The choice of notations and their restrictions allow for an
efficient implementation. This does however depend on the provision of
certain functionality to support the execution of a collection of communicating
parallel processes and, in particular, many-to-one patterns of communication.
These are described first, as an architectural target for the compilation scheme.

\subsection{Architectural target}

The following defines the basic requirements of the proposed language
notations, independent of a specific hardware or software implementation.

\ben

\item {\bf Processor addressing}.
Each processor in a system of $p$ processors has a unique integer ID in the
range $0$ to $p-1$ identifying it.

\item {\bf Multi-threading}.
A processor has the ability to support multiple concurrent \emph{threads} of
execution and any thread has the ability to create additional threads.

\item {\bf Point-to-point communication}.
Any two threads can communicate by passing messages over bidirectional
point-to-point \emph{channels}. A channel is composed of two \emph{channel
ends} that are each local to a thread. A channel end has an ID that combines a
local unique ID with the processor's ID so that it can be uniquely identified
in a system.
Before a process $p$ can send a message to another process $q$, it must set
the destination of a local channel end to be the channel end ID of $q$, that
$q$ is using to receive messages. It is not necessary for $q$ to specify
$p$ as the source unless it sends a message in return to $p$.
All messages are delivered in-order.

\item {\bf Many-to-one connections}.
A channel end may be specified as a destination by multiple senders. In this
case, a sender must be able to establish a connection to ensure other messages
from different senders cannot be delivered and interrupt a communication
sequence.

\een

These requirements are based on the INMOS transputer~\cite{Transputer88} and
related XMOS XS1~\cite{XS1} architectures, which provide low-level or hardware
support for them.
Other larger-scale message passing architectures such as
BlueGene/L~\cite{BlueGeneLMsg05} and BlueGene/Q~\cite{BlueGeneQMsg11} realise
similar concepts in their software point-to-point messaging layer.

\subsection{Scheme}

\subsubsection{Compile-time process allocation}

\begin{figure}[t]

%\tikzset{/tikz/external/export=true}%
%\tikzsetnextfilename{#1}%
%\tikzpicturedependsonfile{#1.tex}%
%\tikzsetnextfilename{#1}%
\begin{center}%
\input{figures/processor-allocation}%
\end{center}%
%\tikzset{/tikz/external/export=false}

\caption{Illustration of the runtime use of processors according to the
compile-time allocation for \proc{allocation-example}, `step' relates to
the sequence of execution.}
\label{fig:processor-allocation}

\end{figure}
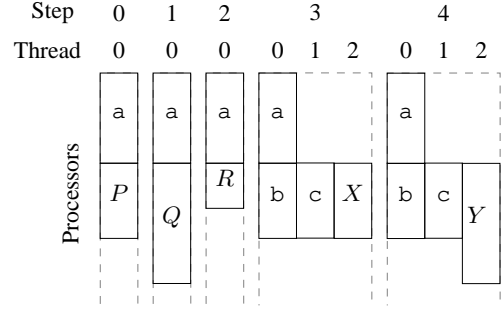

As the size of all process arrays (both replicated processes and servers) can
be determined at compile-time, it is possible to determine a complete static
schedule for the allocation of processes to processors.
This maps process arrays to contiguous blocks of processors and logically
adjacent processes to the same processor.
% Example
For example, the runtime use (and reuse) of processors by
\proc{allocation-example} is illustrated by \fig{processor-allocation}. This
dynamic behaviour is analogous to the allocation of stack frames in memory for
procedure calls.
\begin{Process}
\begin{code*}
\w{server} a \w{is} A(\(\cdots\))[\(n\)] &
\{ \(P\); \(Q\); \(R\);
  \{ \w{server} b \w{is} B(\(\cdots\))[\(m\)] |
    \w{server} c \w{is} C(\(\cdots\))[\(m\)] |
    \{ \(X\); \(Y\) \}
  \}
\}
\end{code*}
\label{proc:allocation-example}
\end{Process}

% Algorithm for processors
Allocation is performed by initialising a \emph{base processor} $b$ to be ID 0. A
process is then assigned to processor $b$ and for each distributed parallel
composition that it contains, $n$ component processes of it are assigned to
processors $b, b+1, \cdots, b+n-1$. The allocation is then applied recursively
to each component process with $b$ set to $b+n$.
%
% Threads
Parallel composition with local distribution is compiled into thread-based
execution with instruction sequences to perform initialisation, start execution
and synchronise before termination.

\subsubsection{Server communication}
% calls on a single channel to message passing sequences, blocking others
% avoiding deadlock with servers (many-to-one connections)
% addressing and connection to servers

% Addressing and connecting
A single server is addressed by its processor ID and local channel end ID. This
can be packed into a single word and passed as a reference. An array of
servers are addressed by a base processor ID, common local channel end ID and
an offset. This allows a normal server call $s$.$c(\cdots)$ or subscripted call
$s$[$i$].$c(\cdots)$, where $s$ is the server reference, to exactly specify a
particular server.

% Calls
The set of calls for a server are implemented with this single channel and each
call is assigned an ID unique to the server.  
%
% Call compilation
Let $a_0, a_1, \cdots, a_{n-1}$ be a set of actual parameters and $P$ be a
process making a call $c$ to a server $s$ of the form \code{$s$.$c$($a_0, a_1,
\cdots, a_{n-1}$).} Then, for a channel end $c$ local to $P$, it is compiled as
the sequence: 
\ben

\item set the destination of $c$ to be the channel end of $s$;

\item connect to $s$;

\item send the channel end ID for $c$;

\item send the call ID for $s$;

\item send each actual parameter $a_i$;

\item receive each referenced actual $a_i'$ and set $a_i\gets a_i'$;

\item disconnect from $s$.

\een
Once the client has connected to the server, it sends the identity of its
channel end so that the server can make the necessary corresponding responses
to the above sequence.
%
% Many-to-one access
By establishing a connection with the server, calls made by other clients will
block until the server becomes free. In this sense, server calls are atomic.

% Avoiding deadlock
A key issue in the implementation of servers is to guarantee that calls always
complete. In a simple implementation, there is potential for deadlock to occur.
This is caused by a situation where multiple clients are waiting for a busy
server. If to service a call the server must perform communication, it might
not be possible to establish a route in the network due to waiting requests
holding network resources.
To avoid this, a server must always be able to consume requests so that a call
is always guaranteed to complete. In practice, the number of clients accessing
any one server is likely to be small and a small queue, with a size
logarithmically related to the number of clients, will probably suffice for most
programs. To avoid deadlock when the queue becomes full, clients can reattempt
to connect, at a rate according to an exponential backoff, similar to the
Ethernet protocol.
Alternatively, two separate physical or logically partitioned networks could be
used, one for server calls and the other for general communication. This way,
queued calls would never interfere with any external communication a server
makes.

% Efficiency in general
%With regards to the efficiency of server communication, with an implementation
%that can provide low level support for communication, the overhead would be
%comparable to any other distributed mechanism such as a cache.

\subsubsection{Process distribution}
% static/dynamic process distribution

The processor allocation is known for each process at compile-time. At run
time, the instruction sequence constituting a process must be available at a
processor that is scheduled to execute it. There are two approaches that can be
taken to this.
With \emph{static distribution}, compilation would produce a set of $p$ binary
images for a $p$ processor system, with each binary containing all the
processes that will be executed by the given processor. This requires each
processor to have a large enough memory to store every process that it will
execute over the course of a program, in addition to the memory requirements of
each process. For large $p$, the size of the binary package could also be
significant.
With \emph{dynamic distribution}, processes are loaded onto processors at
runtime, before they are executed. Compilation produces two binaries, a
\emph{master} image that contains all the program and a \emph{slave} image that
waits to receive processes to execute. The benefit of this is a smaller
per-processor memory requirement and binary package independent of the size of
a system. Dynamic distribution can be made efficient by employing
recursion~\cite{Hanlon11}.
% TODO:? over an XMOS platform in several microseconds, comparable to a
% procedure call...

% Process context
In addition to a component parallel process being available at a processor,
execution on a remote processor also requires the complete lexical environment,
i.e.\ all of the variables it uses that are external to its scope. This can be
determined at compile-time and message passing sequences generated both to
supply these variables and to receive any updates to them when the process
terminates. 

\section{Examples\label{sec:examples}}

This section presents three example programs to demonstrate the proposed
notations: matrix multiplication, a ray tracer and a compiler. The choice of
these is based on general-purpose applications that require different styles of
parallelism.

\subsection{Matrix multiplication\label{sec:matrix-multiply}}

Matrix multiplication is widely used in scientific programs. It is inherently
data-parallel and the most scalable parallel formulations employ message
passing structures.  Cannon's algorithm~\cite{Cannon69} is a simple distributed
algorithm that is structured as a 2D grid.

\begin{figure*}[t]
\centering
\subfloat[Load phase]{\input{figures/multiply-load}%
\label{fig:multiply-load}}%
\hfil
\subfloat[Multiply phase]{\input{figures/multiply}%
\label{fig:multiply-comp}}

\caption{Process distribution and communication structures for successive
  phases of the matrix multiply program. Each employs different communication
  structures; loading performs a sequence of calls to the server array and the
  multiplication algorithm performs only local server accesses, but with
grid-based message passing communication.}

\label{fig:multiply-phases}
\end{figure*}
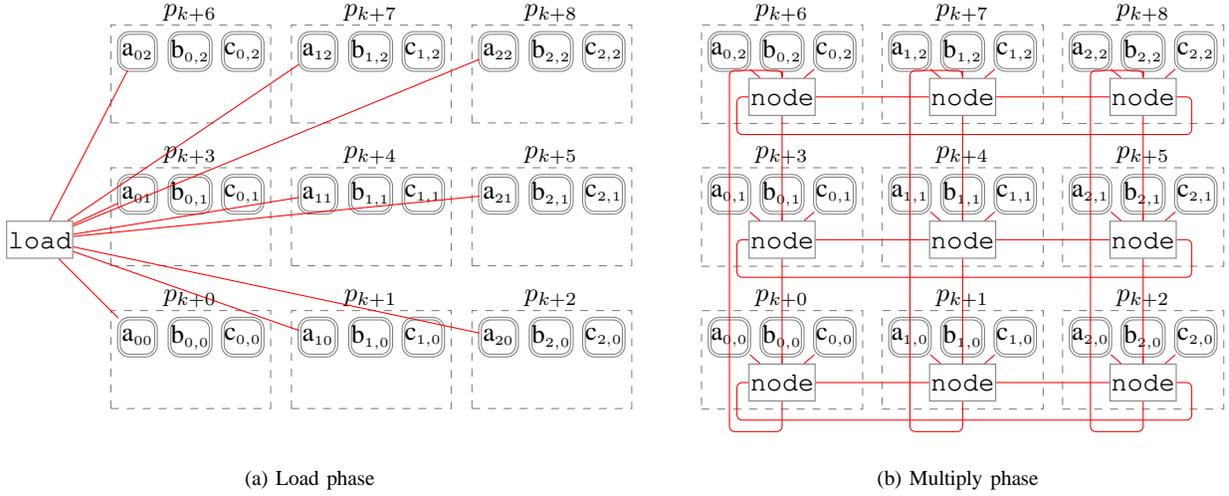

For an $n\times n$ grid of processes, this can be expressed as
\proc{multiply-comp}.  It takes three arrays of sub-matrix servers (\ttt{a},
\ttt{b} and \ttt{c}) as parameters that represent the input and result
matrices. The subroutine proceeds by creating a 2D array of nodes with each
node connected by channels in four directions and assigned a single sub-matrix
server.  The \ttt{node} process performs computations on the local sub matrices
sends and receives sub-matrices in each direction according to the algorithm.
This subroutine encapsulates the algorithm, separating the message passing
implementation from the distributed representation of the matrices. The layout
of this is illustrated in \fig{multiply-comp}. 

% TODO: encapsulation of process structure -- separation from server array

% Gloabal input and output
A subroutine like this will most likely be employed as a component of a
more complex program, but even included in a program that does nothing else, it
requires additional components for the initialisation of the input matrices and
a way to output the result. A simple way to do this is to directly read or
write values to the distributed matrices in a global initialisation phase.
\proc{multiply-load}, for example, iterates over each sub matrix and performs
initialisation directly. A similar process could be conducted to output the
result.
A complete minimal program to perform matrix multiplication could then be
composed as \proc{multiply-program} where the three matrices are declared as
server arrays with a layered distribution. The client process sequentially
loads the input matrices, performs the multiplication and outputs the result.
\fig{multiply-phases} illustrates the distribution of processes and
communication patterns for the load and multiply phases of the algorithm. 
\begin{Process}
\begin{code*}
\w{proc} multiply(
    \w{server} Matrix[n][n] a,
    \w{server} Matrix[n][n] b, 
    \w{server} Matrix[n][n] c, \w{val} n) \w{is}
\{ \w{chan}[n][n+1] h;
  \w{chan}[n][n+1] v;
  \w{var} x, y;
  \w{par} y = 0 \w{for} n \w{do}
    \w{par} x = 0 \w{for} n \w{do}
      node(a[x][y], b[x][y], c[x][y],
        v[x][y], v[x][(y+1) rem n], 
        h[y][x], h[y][(x+1) rem n])
\}
\end{code*}
\label{proc:multiply-comp}
\end{Process}
\begin{Process}
\begin{code*}
\w{proc} loadMatrix(
    \w{server} Matrix[n][n] m, \w{val} n) \w{is}
\{ \w{var} i, j;
  \w{seq} i=0 \w{for} n \w{do}
    \w{seq} j=0 \w{for} n \w{do}
      loadSubMatrix(m[i][j])
\}
\end{code*}
\label{proc:multiply-load}
\end{Process}
\begin{Process}
\begin{code*}
\w{server} a \w{is} Matrix(M, M)[n][n] |
\w{server} b \w{is} Matrix(M, M)[n][n] |
\w{server} c \w{is} Matrix(M, M)[n][n] |
\{ loadMatrix(a, n);
  loadMatrix(b, n);
  multiply(a, b, c, n);
  output(c, n)
\}
\end{code*}
\label{proc:multiply-program}
\end{Process}

\subsection{Ray tracing}

Ray tracing is a technique for generating realistic 2D images from 3D scenes.
It is highly parallelisable as the calculation of each pixel, based on
intersecting a ray with a world model, can be performed independently. When
the world model is small enough to fit into the memory of a single
processor, a parallel scheme requires only the communication of work and
results. 
When it is larger than a single memory, it has to be distributed and accessible
by all processes calculating ray intersections.

\begin{figure}[t]

%\tikzset{/tikz/external/export=true}%
%\tikzsetnextfilename{#1}%
%\tikzpicturedependsonfile{#1.tex}%
%\tikzsetnextfilename{#1}%
\begin{center}%
\input{figures/ray-tracer}%
\end{center}%
%\tikzset{/tikz/external/export=false}

\caption{Structure of the parallel ray tracer where a world model is provided
by an array of servers and accessed concurrently by a collection of workers.
These are delegated work by a single master process. The world model is
disjointedly distributed ($\ell \geq k+n$) but it could also be layered with the
workers ($\ell=k$).}

\label{fig:ray-tracer}
\end{figure}
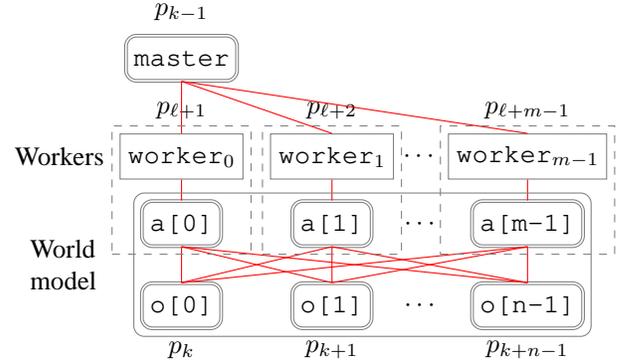

% Overall implementation
A distributed world model has a simple form with the same structure as
the shared memory in \proc{shared-memory}.  Work is distributed in a
\emph{task farm} structure, by a \emph{master} process to a collection of
\emph{worker} processes. 
This is outlined in \proc{ray-tracer} and illustrated in \fig{ray-tracer}.
\proc{ray-tracer} includes separate initialisation and output phases, similar to
the ones described for the matrix multiply program (\proc{multiply-load}). 
\begin{Process}
\begin{code*}
\w{server} master \w{is} Master() &
\w{server} objs \w{is} ObjectStore()[\(n\)] &
\{ \w{server} access \w{is} WorldAccess(objs)[\(m\)] |
  \{ \w{var} i;
    loadWorldModel(access);
    \w{par} i=0 \w{for} \(m\) \w{do}
      worker(master, access);
    output(master)
  \}
\}
\end{code*}
\label{proc:ray-tracer}
\end{Process}

% Workers and access
Each of the $m$ workers can access the world model (distributed over $n$
servers) via a specific server and will do so frequently during the
computation. 
In addition to optimising the implementation of shared memory, it is necessary
to reduce the number and latency of accesses to obtain a scalable ray tracing
algorithm~\cite{Reinhard97}. To do this, each access server can maintain a
summary structure, usually a bounding volume hierarchy (BVH), to minimise
ray-object intersection tests; it can also cache objects.
With existing parallel programming models, this functionality would be
implemented as part of the worker, but in \proc{ray-tracer} it is encapsulated
in the representation of the data, allowing a simple world model interface to
be presented to the workers.

\subsection{Compiler}

%``Compilers are non-trivial programs and mapping a compiler to a parallel
%system provides a non-trivial test case for the problems encountered when
%developing realistic applications for a parallel system''. (Parallel
%compilation for a parallel machine, 1989).

%There's no need to translate and optimise every function in sequence; they can
%all be done at once as soon as the tree is there. There will be some concurrent
%access to parts of the shared tree and symbol table. 
%
%Of course, if the language we are compiling is all about expressing programs as
%collections of independent communicating processes, there will be very little
%concurrent access to the shared tree and symbol table when compiling them. The
%only things in common will be the names associated with the channels and the
%constants used in the protocols

Compilers are complex programs that employ many different algorithmic
techniques and data structures. This makes them a canonical example of a
general-purpose piece of software and a non-trivial test case for mapping
realistic sequential applications to a parallel architecture.  Due to this,
there has been little work on parallel compilation, although there are
opportunities to, particularly during the optimisation and code generation
phases~\cite{Gross89}. In particular, many optimisations can be applied locally
at an expression, statement, block or procedure level, and hence may be
performed independently and in parallel over different parts of a parse tree
or intermediate representation.

The structure of a simple compiler is given in \proc{compiler}. 
\begin{Process}
\begin{code*}
\w{server} store \w{is} TreeStore()[\(n\)] &
\w{server} tree \w{is} TreeAccess(store)[\(m\)] &
\w{server} symbols \w{is} Table() &
\{ parse(tree[0], symbols);
  semantics(tree[0], symbols, \(m\));
  optimise(tree, symbols);
  \{ \w{server} store \w{is} BufStore()[\(l\)] |
    \w{server} buffer \w{is} BufAccess(store) |
    generateInsts(tree[0], buffer);
  \}
\}
\end{code*}
%\caption{}
\label{proc:compiler}
\end{Process}
Two server arrays \ttt{store} and \ttt{tree} provide a concurrently accessible
parse tree, using the same principle as \proc{shared-memory}. Initially, parsing and
semantic analysis phases operate sequentially on the parse tree, using a single
access server. Local optimisations, as part of the \ttt{optimise} subroutine,
can be performed in parallel on the parse tree and this will also require
concurrent access to the symbol table. Finally, instructions are
output sequentially to a distributed buffer. This buffer is declared in a separate
scope to demonstrate it could be included as part of the \ttt{generateInsts}
subroutine.

%\section{Discussion\label{sec:discussion}}
%
%% * Variable sized process arrays (dyanmic distribution/allocation - innefficient)
%
%
%% * Sleeping servers fits nicely with multi-threaded archs & ones that can
%% power down cores when idle
%As servers are only active in response to calls, they could allow or trigger
%the underlying processor to move into a low power state when they are idle.
%
%% * Ability to optimise notiations as they have proper semantics
%
%
%% * Requirements of the capability of the interconnect to support arbitrary
%%   process structures and a simple processor allocation scheme
%
%% Syntax open

\section{Conclusion\label{sec:conclusion}}

% Benefits of the proposal
This paper proposes a simple programming model for expressing scalable parallel
programs. 
A server construct can be used in combination with notations for expressing
local and distributed parallelism to build abstractions for distributed data
structures with both shared and distributed access structures. This gives the
programmer the flexibility to move between shared and distributed forms of data
parallelism, depending on the structure of the program and scalability
requirements. 
Server-based data structures can be composed with other program components in a
similar way to conventional variable declarations and have similar scoping
rules. This allows them to be operated on by sequences of potentially parallel
subroutines, simplifying the task of developing a complex parallel program.

% Implementation highlights
The distribution model allows a compile-time allocation of processing
resources, to produce a static schedule. This provides efficient runtime
performance and predictable timing, which are essential for building programs
that scale to large numbers of cores.
%
%Connection to servers and communication with them can be peformed with a low
%overhead. With the provision of efficient communication mechanisms in the
%hardware, performance of server communication should be comparable with any
%other distributed communication mechanisms, such as the interconnect with a
%cache.
%
The compilation scheme requires support from the architecture, in particular to
provide bounded low latency communications, to support the distribution model
and general patterns of communication between program components and servers,
and in message passing structures such as pipelines, grids and trees.

% Examples highlights
The example programs demonstrate how the proposed notations can be used to
compose computational components that require varied forms of parallelism with
distributed data structures, in a clear and concise way.

\section*{Acknowledgement}

This work was funded by EPSRC grant SB1933.

\bibliographystyle{IEEEtran}
%\def\IEEEbibitemsep{2pt}
\bibliography{IEEEabrv,refs.bib}
%\begin{thebibliography}{1}
%\end{thebibliography}

\end{document}

%% file: preamble.tex
\usepackage{
  amsmath, 
  amssymb, 
  amsthm, 
  amsfonts,
  latexsym,
  graphicx,
  comment,
  url,
  setspace,
  tikz,
  alltt,
}
\usepackage{hyperref}

\usetikzlibrary{calc}
\newcommand{\ben}{\begin{enumerate}}
\newcommand{\een}{\end{enumerate}}
\newcommand{\bds}{\begin{description}}
\newcommand{\eds}{\end{description}}
\newcommand{\bit}{\begin{itemize}}
\newcommand{\eit}{\end{itemize}}
\newcommand{\bci}{\begin{compactitem}}
\newcommand{\eci}{\end{compactitem}}
\newcommand{\bvb}{\begin{verbatim}}
\newcommand{\evb}{\end{verbatim}}
\newcommand{\ttt}[1]{\texttt{#1}}

\newcommand{\fig}[1]{Fig.~\ref{fig:#1}}

\newcommand{\sect}[1]{Section~\ref{sec:#1}}

\newcommand{\proc}[1]{Process~\ref{proc:#1}}

\newtheorem{theorem}{Theorem}

\newtheoremstyle{procstyle} % name
    {0.3em}                    % Space above
    {0.3em}                    % Space below
    {\itshape}                   % Body font
    {}                           % Indent amount
    {\footnotesize\scshape}      % Theorem head font
    {\newline}                   % Punctuation after theorem head
    {\linewidth}                 % Space after theorem head
    {}  % Theorem head spec (can be left empty, meaning ‘normal’)
\usepackage{float}
\floatstyle{plain}
\newfloat{procfloat}{thp}{lop}
\floatname{procfloat}{}
\newcounter{process}
\newenvironment{Process}[1][]{%
\begin{procfloat}[ht]%
\vspace{-1.5mm}%
\refstepcounter{process}%
{\bf Process~\theprocess} #1}
{%\end{minipage}
\vspace{-3.5mm}%
\end{procfloat}}

\newcommand{\w}[1]{{\bf #1}}

\def\codespacing{1.5mm}

\newenvironment{myquote}
{\list{}{\leftmargin=4mm\rightmargin=4mm}\item[]}
{\endlist}

\newenvironment{code*}%
{\vspace{\codespacing}%
\begin{myquote}%
\begin{minipage}{\linewidth}%
\begin{alltt}}
{\end{alltt}%
\end{minipage}
\end{myquote}%
\vspace{\codespacing}
}

\newcommand{\code}[1]{
\vspace{\codespacing}%
\begin{myquote}%
\begin{alltt}%
#1%
\end{alltt}%
\end{myquote}%
\vspace{\codespacing}
}

%% file: figures/par-composition-local.tex
\begin{tikzpicture}[every node/.style={draw=gray, inner sep=1pt, minimum height=0.6cm, minimum width=0.6cm}]

\draw[fill] (0,0) node[rectangle] (P) {$P$}
	++(0.8,0) node[rectangle,label={[below] $p_{k}$}] (Q) {$Q$}
	++(0.8,0) node[rectangle] (R) {$R$};

\draw[dashed,color=gray]
	let
		\p1=(P.north west),
		\p2=(R.north east),
		\p3=(R.south east)
	in
		($(\x1,\y3)+(-0.1,-0.1)$)
		rectangle
		($(\x2,\y1)+(0.1,0.1)$) {};

\end{tikzpicture}

%% file: figures/par-composition.tex
\begin{tikzpicture}[every node/.style={draw=gray, inner sep=1pt, minimum height=0.6cm, minimum width=0.6cm}]

\draw[fill] (0,0) node[rectangle,label={[below] $p_k$}] {$P$}
	++(1,0) node[rectangle,label={[below] $p_{k+1}$}] {$Q$}
	++(1,0) node[rectangle,label={[below]$p_{k+2}$}] {$R$};

\end{tikzpicture}

%% file: figures/par-replication-layered.tex
\def\increment{}
\begin{tikzpicture}[
every node/.style={draw=gray, inner sep=3pt},
labels/.style={above=0.1cm}
]

\draw[fill] (0,0) node[rectangle,label={[labels] $p_k$}] (s0) {$P(1)$}
	++(1.3,0) node[rectangle,label={[labels] $p_{k+1}$}] (s1) {$P(2)$}
	++(1.8,0) node[rectangle,label={[labels] $p_{k+n-1}$}] (sn) {$P(n)$};

% Clients
\draw[fill] ($(s0.south)-(0,0.4)$) node[rectangle] (c0) {$Q(1)$};
\draw[fill] ($(s1.south)-(0,0.4)$) node[rectangle] (c1) {$Q(2)$};
\draw[fill] ($(sn.south)-(0,0.4)$) node[rectangle] (cn) {$Q(n)$};

% Dots
\draw let \p0=(s1.east), \p1=(sn.west) in ($(\x0,\y0)+0.5*(\x1-\x0,0)$)  
	node[draw=none] {$\cdots$};
\draw let \p0=(c1.east), \p1=(cn.west) in ($(\x0,\y0)+0.5*(\x1-\x0,0)$)  
	node[draw=none] {$\cdots$};

% Processor boxes
\foreach \a / \b in {s0/c0,s1/c1,sn/cn} {
	\draw[dashed,color=gray]
	let
		\p1=(\a.north west),
		\p2=(\a.north east),
		\p3=(\b.south east)
	in
		($(\x1,\y3)+(-0.1,-0.1)$)
		rectangle
		($(\x2,\y1)+(0.1,0.1)$) {};
}

\end{tikzpicture}

%% file: figures/par-replication-disjoint.tex
\begin{tikzpicture}[
every node/.style={draw=gray, inner sep=3pt},
labels/.style={above=0cm}
]

\draw[fill] (0,0) node[rectangle,label={[labels] $p_k$}] (p1) {$P(1)$}
	++(1.4,0) node[rectangle,label={[labels] $p_{k+1}$}] (p2) {$P(2)$}
	++(1.6,0) node[rectangle,label={[labels] $p_{k+n-1}$}] (pn) {$P(n)$};

\draw[fill] (0,-1.1) node[rectangle,label={[labels] $p_{\ell}$}] (q1) {$Q(1)$}
	++(1.4,0) node[rectangle,label={[labels] $p_{\ell+1}$}] (q2) {$Q(2)$}
	++(1.6,0) node[rectangle,label={[labels] $p_{\ell+m-1}$}] (qn) {$Q(m)$};

% Dots
\draw ($(p2.east)!0.5!(pn.west)$) node[draw=none] {$\cdots$};
\draw ($(q2.east)!0.5!(qn.west)$) node[draw=none] {$\cdots$};

\end{tikzpicture}

%% file: figures/server.tex
\def\increment{}
\begin{tikzpicture}[
every node/.style={draw=gray, inner sep=3pt},
labels/.style={above=0.1cm},
link/.style={color=red},
server/.style={rectangle,double,rounded corners},
]

\draw 
	(0,0) node[label={[above=4] Server}] (state) {state}
	++(0, -0.3) node[draw=none,anchor=north, text width=1cm, text centered] (calls) {call$_1$\\call$_2$\\$\vdots$\\call$_n$};

\draw[server,color=gray]
	let
		\p1=(state.north west),
		\p2=(calls.north east),
		\p3=(calls.south east),
		\p4=(calls.south west)
	in
		($(\x4,\y3)+(-0.1,-0.1)$)
		rectangle
		($(\x2,\y1)+(0.1,0.1)$) {};

\draw[fill] (1.8,-0.1) node[rectangle] (c1) {client$_1$}
++(0,-0.7) node[rectangle] (c2) {client$_2$}
++(0,-1.2) node[rectangle] (cm) {client$_m$};

% Dots
\draw let \p0=(c2.south), \p1=(cm.north) in 
	($(\x0,\y0)-0.5*(0,\y0-\y1)$)  node[draw=none,text height=10] {$\vdots$};

\draw[link,->] (c1.west) -- (calls);
\draw[link,->] (c2.west) -- (calls);
\draw[link,->] (cm.west) -- (calls);

\end{tikzpicture}

%% file: figures/server-replicated.tex
\begin{tikzpicture}[
every node/.style={draw=gray, inner sep=3pt, 
	minimum height=0.6cm, minimum width=1cm},
labels/.style={above},
server/.style={rectangle,double,rounded corners},
]

% Processes
\draw[fill] (0,0) node[server,label={[labels] $p_k$}] (s0) {\texttt{s[0]}}
	++(1.2,0) node[server,label={[labels] $p_{k+1}$}] (s1) {\texttt{s[1]}}
	++(2,0) node[server,label={[labels] $p_{k+n-1}$}] (sn) {\texttt{s[n-1]}};

% Dots
\draw let \p0=(s1.east), \p1=(sn.west) in ($(\x0,\y0)+0.5*(\x1-\x0,\y1)$)  
	node[draw=none] {$\cdots$};

\end{tikzpicture}

%% file: figures/server-distributed-shared-memory.tex
\begin{tikzpicture}[
every node/.style={draw=gray, inner sep=3pt, 
	minimum height=0.6cm, minimum width=1cm},
tlabels/.style={above},
blabels/.style={below=1.2cm},
link/.style={
color=red
},
labels/.style={above=0cm},
server/.style={rectangle,double,rounded corners},
]

% Stores
\draw[fill] (0,0) node[server,label={[blabels] $p_k$}] (s0) {\texttt{s[0]}}
	++(1.5,0) node[server,label={[blabels] $p_{k+1}$}]  (s1) {\texttt{s[1]}}
	++(2,0) node[server,label={[blabels] $p_{k+n-1}$}] (sn) {\texttt{s[n-1]}};

% Access
\draw[fill] ($(s0.south)+(0,1.4)$) node[server] (a0) {\texttt{a[0]}};
\draw[fill] ($(s1.south)+(0,1.4)$) node[server] (a1)  {\texttt{a[1]}};
\draw[fill] ($(sn.south)+(0,1.4)$) node[server] (an) {\texttt{a[m-1]}};

% Clients
\draw[fill] ($(a0.south)+(0,1.2)$) node[rectangle,label={[tlabels] $p_{\ell}$}] (c0) {};
\draw[fill] ($(a1.south)+(0,1.2)$) node[rectangle,label={[tlabels] $p_{\ell+1}$}] (c1) {};
\draw[fill] ($(an.south)+(0,1.2)$) node[rectangle,label={[tlabels] $p_{\ell+m-1}$}] (cn) {};

% Dots
\draw ($(s1.east)!0.5!(sn.west)$) node[draw=none] {$\cdots$};
\draw ($(a1.east)!0.5!(an.west)$) node[draw=none] {$\cdots$};
\draw ($(c1.east)!0.5!(cn.west)$) node[draw=none] {$\cdots$};

% Labels
\node[draw=none] (l1) at ($(s0.west)-(1,0)$) {Storage};
\node[draw=none] (l1) at ($(a0.west)-(1,0)$) {Access};
\node[draw=none] (l1) at ($(c0.west)-(1,0)$) {Clients};

% access-store links
\foreach \x in {a0,a1,an}
	\foreach \y in {s0,s1,sn}
		\draw [link] (\x.south) -- (\y.north);

% Worker-access links
\foreach \x in {0,1,n}
	 \draw [link] (c\x.south) -- (a\x.north);

% Processor boxes
\foreach \a / \b in {c0/a0,c1/a1,cn/an} {
	\draw[dashed,color=gray]
	let
		\p1=(\a.north west),
		\p2=(\a.north east),
		\p3=(\b.south east),
		\p4=(\b.south west)
	in ($(\x4,\y3)+(-0.1,-0.1)$) rectangle ($(\x3,\y2)+(0.1,0.1)$) {};
}

\end{tikzpicture}

%% file: figures/server-distributed-memory.tex
\usetikzlibrary{calc}
\begin{tikzpicture}[
every node/.style={draw=gray, inner sep=3pt, 
	minimum height=0.6cm, minimum width=1cm},
labels/.style={above},
link/.style={color=red},
server/.style={rectangle,double,rounded corners},
]

% Servers
\draw[fill] (0,0) node[server] (s0) {\texttt{s[0]}}
	++(1.5,0) node[server]  (s1) {\texttt{s[1]}}
	++(2.2,0) node[server] (sn) {\texttt{s[n-1]}};

% Clients
\draw[fill] ($(s0.north)+(0,0.6)$) node[rectangle,label={[labels] $p_k$}] (p0) {};
\draw[fill] ($(s1.north)+(0,0.6)$) node[rectangle,label={[labels] $p_{k+1}$}] (p1) {};
\draw[fill] ($(sn.north)+(0,0.6)$) node[rectangle,label={[labels] $p_{k+n-1}$}] (pn) {};

% Labels
\node[draw=none] (l1) at ($(s0.west)-(1,0)$) {Servers};
\node[draw=none] (l1) at ($(p0.west)-(1,0)$) {Clients};

% Dots
\draw let \p0=(s1.east), \p1=(sn.west) in ($(\x0,\y0)+0.5*(\x1-\x0,0)$)  
	node[draw=none] {$\cdots$};
\draw let \p0=(p1.east), \p1=(pn.west) in ($(\x0,\y0)+0.5*(\x1-\x0,0)$)  
	node[draw=none] {$\cdots$};

% Processor boxes
\foreach \a / \b in {p0/s0,p1/s1,pn/sn} {
	\draw[dashed,color=gray]
	let
		\p1=(\a.north west),
		\p2=(\a.north east),
		\p3=(\b.south east),
		\p4=(\b.south west)
	in
		($(\x4,\y3)+(-0.1,-0.1)$)
		rectangle
		($(\x3,\y1)+(0.1,0.1)$) {};
}

% Communication links
\foreach \x / \y in {s0/p0,s1/p1,sn/pn}
  \draw [link] (\x.north) -- (\y.south);

\end{tikzpicture}

%% file: figures/processor-allocation.tex
\usetikzlibrary{positioning,fit,chains}
\usetikzlibrary{backgrounds}
\begin{tikzpicture}[
every node/.style={draw, inner sep=0, minimum height=0.6cm, minimum width=0.5cm, text centered},
labels/.style={above},
box/.style={draw,line width=0.1pt,align=center},
bbox/.style={dashed,draw,line width=0.1pt,align=center,color=gray},
container/.style={dashed,line width=0.1pt,color=gray},
add reference/.style={insert path={%
    coordinate [pos=0,xshift=-0.5\pgflinewidth,yshift=-0.5\pgflinewidth] (#1 south west) 
    coordinate [pos=1,xshift=0.5\pgflinewidth,yshift=0.5\pgflinewidth]   (#1 north east)
    coordinate [pos=.5] (#1 center)                        
    (#1 south west |- #1 north east)     coordinate (#1 north west)
    (#1 center     |- #1 north east)     coordinate (#1 north)
    (#1 center     |- #1 south west)     coordinate (#1 south)
    (#1 south west -| #1 north east)     coordinate (#1 south east)
    (#1 center     -| #1 south west)     coordinate (#1 west)
    (#1 center     -| #1 north east)     coordinate (#1 east)   
}}
]

\small
\newcommand\vertspace[1]{\vrule height #1 depth #1 width 0pt}
\def\spacing{0.2}
\def\height{2.6cm}

% 0
\begin{scope}[start chain=going below ,box,node distance=-0.1pt]
\draw node[on chain] (0) at (0,0) 
{\vertspace{0.6cm}\texttt{a}};
\node[on chain] (p) {
\vertspace{0.5cm}$P$};
\begin{scope}[on background layer]
\draw[container] let \p0=(0.north west), \p1=(p.south east) in
	($(\x0,\y0)+(0.1pt,0pt)$) rectangle (\x1,-\height) [add reference=container0];
\end{scope}
\draw node[draw=none, above=0.5] (phaseLabel) at (container0 south) {0};
\draw node[draw=none, above=0] (threadLabel) at (container0 south) {0};
\end{scope}

% 1
\begin{scope}[start chain=going below,box,node distance=-0.1pt]
\draw node[on chain,anchor=north west] (1) at ($(0.north east)+(\spacing,0)$) 
{\vertspace{0.6cm}\texttt{a}};
\node[on chain] (q){
\vertspace{0.8cm}
$Q$};
\begin{scope}[on background layer]
\draw[container] let \p0=(1.north west), \p1=(q.south east) in 
	($(\x0,\y0)+(0.1pt,0pt)$) rectangle (\x1,-\height) [add reference=container1];
\end{scope}
\draw node[draw=none, above=0.5] at (container1 south) {1};
\draw node[draw=none, above=0] at (container1 south) {0};
\end{scope}

% 2
\begin{scope}[start chain=going below,box,node distance=-0.1pt]
\draw node[on chain,anchor=north west] (2) at ($(1.north east)+(\spacing,0)$) 
{\vertspace{0.6cm}\texttt{a}};
\node[on chain] (r){
\vertspace{0.3cm}
$R$};
 \begin{scope}[on background layer]
\draw[container] let \p0=(2.north west), \p1=(r.south east) in 
	($(\x0,\y0)+(0.1pt,0pt)$) rectangle (\x1,-\height) [add reference=container2];
\end{scope}
\draw node[draw=none, above=0.5] at (container2 south) {2};
\draw node[draw=none, above=0] at (container2 south) {0};
\end{scope}

% 3
\begin{scope}

\draw node[box,anchor=north west] (3) at ($(2.north east)+(\spacing,0)$) 
{\vertspace{0.6cm}\texttt{a}};

\node[box,anchor=north] (bs) at ($(3.south)-(0,-0.1pt)$) 
{\vertspace{0.5cm}\texttt{b}};

\node[box,anchor=west] (cs) at ($(bs.east)-(0.1pt,-0.1pt)$) 
{\vertspace{0.5cm}\texttt{c}};

\node[box,anchor=north west] (xs) at ($(cs.north east)-(0.1pt,0.1pt)$) 
{\vertspace{0.5cm}$X$};

\begin{scope}[on background layer]
\draw[container] let \p0=(3.north west), \p1=(xs.south east) in 
	($(\x0,\y0)+(0.1pt,0pt)$) rectangle (\x1,-\height) [add reference=container3];
\end{scope}
\draw node[draw=none, above=0.5] at (container3 south) {3};
\draw node[draw=none, above] at ($(container3 south)-(0.5,0)$) {0};
\draw node[draw=none, above] at (container3 south) {1};
\draw node[draw=none, above] at ($(container3 south)+(0.5,0)$) {2};
\end{scope}

% 4
\begin{scope}

\draw node[box,anchor=north west] (4) at ($(container3 south east)+(\spacing,0)$) 
{\vertspace{0.6cm}\texttt{a}};

\node[box,anchor=north] (bs) at ($(4.south)-(0,-0.1pt)$) 
{\vertspace{0.5cm}\texttt{b}};

\node[box,anchor=west] (cs) at ($(bs.east)-(0.1pt,-0.1pt)$) 
{\vertspace{0.5cm}\texttt{c}};

\draw node[box,anchor=north west] (y) at ($(cs.north east)-(0.1pt,0.1pt)$) {
\vertspace{0.8cm}
$Y$
};

\begin{scope}[on background layer]
\draw[container] let \p0=(4.north west), \p1=(y.south east) in 
	($(\x0,\y0)+(0.1pt,0pt)$) rectangle (\x1,-\height) [add reference=container4];
\end{scope}
\draw node[draw=none, above=0.5] at (container4 south) {4};
\draw node[draw=none, above] at ($(container4 south)-(0.5,0)$) {0};
\draw node[draw=none, above] at (container4 south) {1};
\draw node[draw=none, above] at ($(container4 south)+(0.5,0)$) {2};
\end{scope}

% Trim bottom
\draw[color=white, fill] let \p0=(container0 north west), \p1=(container4 north east) in ($(\x0,\y0)-(0,0.01)$) rectangle ($(\x1,\y1)+(0,0.1)$);

% Processors label
\node [left of=container0 west, left=-0.3, rotate=90, anchor=north, draw=none] {Processors};

% Phase label
\node [draw=none] at ($(phaseLabel.west)-(0.63,0)$) {Step};

% Thread label
\node [draw=none] at ($(threadLabel.west)-(0.7,0)$) {Thread};

\end{tikzpicture}

%% file: figures/multiply-load.tex
\usetikzlibrary{calc}
\def\scalingx{2.4}
\def\scalingy{1.9}
\begin{tikzpicture}[
node/.style={rectangle,draw, inner sep=0.5pt,
minimum height=14, minimum width=14, draw=gray
},
link/.style={color=red, rounded corners=2,
	draw opacity=1},
server/.style={rectangle,double,rounded corners,node},
]

% Positioning
\fill[white] (-1,-1) rectangle (5,5) {};

% Cells
\foreach \Y in {0,1,2}
  \foreach \X in {0,1,2} {
		\pgfmathsetmacro\result{\X + (\Y*3)}
		% Client
		\node [draw=none,color=white] (grid-\X-\Y) at (\X*\scalingx,\Y*\scalingy) {node};
		% Servers
 		\node [server,above=10] (a-\X-\Y) at ($(grid-\X-\Y)-(0.7,0)$) {a{\tiny$_{\X\Y}$}};
 		\node [server,above=10] (b-\X-\Y) at (grid-\X-\Y) {b{\tiny$_{\X,\Y}$}};
 		\node [server,above=10] (c-\X-\Y) at ($(grid-\X-\Y)+(0.7,0)$) {c{\tiny$_{\X,\Y}$}};
		\node [above=25] at (grid-\X-\Y) {$p_{k+\pgfmathprintnumber{\result}}$};
		% Links
		%\draw[link] (grid-\X-\Y) -- (a-\X-\Y);
		% Box
		\draw[dashed,color=gray] let
			\p1=(a-\X-\Y.north west),
			\p2=(c-\X-\Y.north east),
			\p3=(grid-\X-\Y.south)
			in
			($(\x1,\y3)+(-0.1,-0.1)$) rectangle ($(\x2,\y1)+(0.1,0.1)$) {};
  }

\draw let \p0=(grid-0-0.south), \p1=(grid-0-2.north) in
	node[node]  at ($(-2,\y0)+0.5*(0,\y1-\y0)$) (master) {\texttt{load}};

% Horizontal channels
%\foreach \y in {0,1,2}
%	\foreach \x / \xn in {0/1,1/2}
%		\draw [link] (grid-\x-\y) -- (grid-\xn-\y);

% Vertical channels
%\draw [link] (grid-2-0) -- (grid-0-1);
%\draw [link] (grid-2-1) -- (grid-0-2);

% Draw master channels
%\draw [link] (master.north) |-  ($(grid-2-2.east)+(0.6,-0.5)$) |- (grid-2-2.east);
%\draw [link] (master.south) |- (grid-0-0);

\foreach \Y in {0,1,2}
	\foreach \X in {0,1,2}
		\draw [link] (master) -- (a-\X-\Y);

\end{tikzpicture}

%% file: figures/multiply.tex
\def\scalingx{2.4}
\def\scalingy{1.9}
\begin{tikzpicture}[
node/.style={rectangle,draw, inner sep=0.5pt,
	minimum height=14, minimum width=14, draw=gray},
link/.style={color=red, rounded corners=2, draw opacity=1},
server/.style={rectangle,double,rounded corners,node},
]

% Positioning
\fill[white] (-1,-1) rectangle (5,5) {};

% Cells
\foreach \Y in {0,1,2}
  \foreach \X in {0,1,2} {
		\pgfmathsetmacro\result{\X + (\Y*3)}
		% Client
		\node [node] (grid-\X-\Y) at (\X*\scalingx,\Y*\scalingy) {\texttt{node}};
		% Servers
 		\node [server,above=10] (a) at ($(grid-\X-\Y)-(0.7,0)$) {a{\tiny$_{\X,\Y}$}};
 		\node [server,above=10] (b) at (grid-\X-\Y) {b{\tiny$_{\X,\Y}$}};
 		\node [server,above=10] (c) at ($(grid-\X-\Y)+(0.7,0)$) {c{\tiny$_{\X,\Y}$}};
		\node [above=25] at (grid-\X-\Y) {$p_{k+\pgfmathprintnumber{\result}}$};
		% Links
		\draw[link] (grid-\X-\Y) -- (a);
		\draw[link] (grid-\X-\Y) -- (b);
		\draw[link] (grid-\X-\Y) -- (c);
		% Box
		\draw[dashed,color=gray] let
			\p1=(a.north west),
			\p2=(c.north east),
			\p3=(grid-\X-\Y.south)
			in
			($(\x1,\y3)+(-0.1,-0.1)$) rectangle ($(\x2,\y1)+(0.1,0.1)$) {};
  }

% Horizontal channels
\foreach \y in {0,1,2}
	\foreach \x / \xn in {0/1,1/2}
		\draw [link] (grid-\x-\y) -- (grid-\xn-\y);

% Vertical channels
\foreach \y / \yn in {0/1,1/2}
	\foreach \x / \xn in {0,1,2}
		\draw [link] (grid-\x-\y) -- (grid-\x-\yn);

% Wrap-around y
\foreach \x in {0,1,2} {
	\draw [link] (grid-\x-0.south) |- ++(-0.2,-0.4) -| ++(-0.5,1) |-
		($(grid-\x-2.north)+(0,0.1)$) |- (grid-\x-2.north);
};

% Wrap-around x
\foreach \y in {0,1,2} {
	\draw [link] (grid-0-\y) -| ++(-0.6,-0.5) -- ($(grid-2-\y.east)+(0.2,-0.5)$)  |- (grid-2-\y.east);
};

\end{tikzpicture}

%% file: figures/ray-tracer.tex
\usetikzlibrary{calc}
\begin{tikzpicture}[
every node/.style={draw=gray, inner sep=3pt, 
	minimum height=0.6cm, minimum width=1cm},
tlabels/.style={above},
blabels/.style={below=1.2cm},
link/.style={
color=red
},
server/.style={rectangle,double,rounded corners},
]

% Object stores
\draw[fill] (0,0) node[server,label={[blabels] $p_k$}] (s0) {\texttt{o[0]}}
	++(2,0) node[server,label={[blabels] $p_{k+1}$}]  (s1) {\texttt{o[1]}}
	++(2.6,0) node[server,label={[blabels] $p_{k+n-1}$}] (sn) {\texttt{o[n-1]}};

% Object access
\draw[fill] ($(s0.south)+(0,1.4)$) node[server] (a0) {\texttt{a[0]}};
\draw[fill] ($(s1.south)+(0,1.4)$) node[server] (a1)  {\texttt{a[1]}};
\draw[fill] ($(sn.south)+(0,1.4)$) node[server] (an) {\texttt{a[m-1]}};

% Workers
\draw[fill] ($(a0.south)+(0,1.2)$) node[rectangle,label={[tlabels] $p_{\ell+1}$}] (w0) {\texttt{worker}$_0$};
\draw[fill] ($(a1.south)+(0,1.2)$) node[rectangle,label={[tlabels] $p_{\ell+2}$}] (w1)  {\texttt{worker}$_1$};
\draw[fill] ($(an.south)+(0,1.2)$) node[rectangle,label={[tlabels] $p_{\ell+m-1}$}] (wn) {\texttt{worker}$_{m-1}$};

% Master
\draw[fill] ($(w0.north)+(0,1)$) node[server,label={[tlabels] $p_{k-1}$}] (master) {\texttt{master}};

% Dots
\draw ($(a1.east)!0.5!(an.west)$)  node[draw=none] {$\cdots$};
\draw ($(w1.east)!0.5!(wn.west)$)  node[draw=none] {$\cdots$};
\draw ($(s1.east)!0.5!(sn.west)$)  node[draw=none] {$\cdots$};

% Labels
\draw ($(s0.south)!0.5!(a0.north)$) node[draw=none,left=0.9cm,text width=1cm] {World\\model};
\node[draw=none] (l2) at ($(w0.west)-(0.8,0)$) {Workers};

% Master-worker links
\foreach \x in {w0,w1,wn}
		\draw [link] (\x.north) -- (master.south);

% access-store links
\foreach \x in {a0,a1,an}
	\foreach \y in {s0,s1,sn}
		\draw [link] (\x.south) -- (\y.north);

% Worker-access links
\foreach \x in {0,1,n}
	 \draw [link] (w\x.south) -- (a\x.north);

% Processor boxes
\foreach \a / \b in {w0/a0,w1/a1,wn/an} {
	\draw[dashed,color=gray]
	let
		\p1=(\a.north west),
		\p2=(\a.north east),
		\p3=(\b.south east)
	in ($(\x1,\y3)+(-0.1,-0.1)$) rectangle ($(\x2,\y1)+(0.1,0.1)$) {};
}

% World model box
\draw[color=gray,rounded corners]
let
	\p1=(a0.north west),
	\p2=(a0.north east),
	\p3=(sn.south east)
in 	($(\x1,\y3)+(-0.1,-0.1)$) rectangle ($(\x3,\y1)+(0.1,0.1)$) {};

\end{tikzpicture}